\documentclass[11pt]{article}
\usepackage{subfigure}
\usepackage{epsf}
\usepackage{psfrag}
\usepackage{cite}
\usepackage{amssymb}
\usepackage{amsmath}
\usepackage{mathrsfs}
\usepackage{graphicx}
\usepackage{latexsym}
\usepackage{array}

\newtheorem{theorem}{Theorem}

\newtheorem{corollary}{Corollary}
\newtheorem{lemma}{Lemma}
\newcommand{\eqdef}{\stackrel{\triangle}{=}}
\newcommand{\bc}{{\bf c}}
\newcommand{\bg}{{\bf g}}
\newcommand{\bG}{{\bf G}}
\newcommand{\bH}{{\bf H}}
\newcommand{\bh}{{\bf h}}

\newcommand{\mC}{{\mathcal C}}

\newcommand{\dmin}{d_{\min}}
\newcommand{\rs}{\hspace*{-0.1in} }
\newcommand{\aline}[1]{\rs&{#1}&\rs}
\newcommand{\GF}{\text{GF}}

\begin{document}
\bibliographystyle{unsrt}
\renewcommand{\baselinestretch}{1.35}\normalsize
\begin{titlepage}

\thispagestyle{empty}

\title{\bf On Expanded Cyclic Codes}

\author{Yingquan Wu \\[0.1in]
Link\_A\_Media Devices Corp. \\
2550 Walsh Ave, Suite 200 \\
Santa Clara, CA 95051, \ USA
}

\end{titlepage}
\maketitle

\begin{abstract}
The paper has a threefold purpose. The first purpose is to present an explicit description of expanded cyclic codes defined in $\GF(q^m)$.
The proposed explicit construction of expanded generator matrix and expanded parity check matrix 
maintains the symbol-wise algebraic structure and thus keeps many important original characteristics.
The second purpose of this paper is to identify a class of constant-weight cyclic codes. Specifically,
we show that a well-known class of $q$-ary BCH codes excluding the all-zero codeword are constant-weight cyclic codes. 
Moreover, we show this class of codes achieve the Plotkin bound.
The last purpose of the paper is to characterize expanded cyclic codes utilizing the proposed expanded generator matrix and parity check matrix. 
We characterize the properties of component codewords of a codeword and particularly identify the precise conditions under which a codeword can be represented by a subbasis. 
Our developments reveal an alternative while more general view on the subspace subcodes of Reed-Solomon codes.
With the new insights, we present an improved lower bound on the minimum distance of an expanded cyclic code 
by exploiting the generalized concatenated structure. 
We also show that the fixed-rate binary expanded Reed-Solomon codes are asymptotically ``bad", 
in the sense that the ratio of minimum distance over code length diminishes with code length going to infinity.
It overturns the prevalent conjecture that they are ``good" codes and deviates from 
the ensemble of generalized Reed-Solomon codes which asymptotically achieves the Gilbert-Varshamov bound.
\end{abstract}

\newpage 

\section{Introduction}

The paper has a threefold purpose, the first of which is to present an explicit description of expanded cyclic codes defined in $\GF(q^m)$.
An explicit construction of an expanded generator matrix is given in terms of trace and dual/complementary basis \cite{Lidl, MacWilliams}.
An expanded parity check matrix can be constructed through replacing each element with its matrix representation \cite{MacWilliams}. 
The proposed explicit construction of expanded generator matrix and expanded parity check matrix 
maintains the symbol-wise algebraic structure and thus keeps many important original properties.

The second purpose of this paper is to identify a class of constant-weight cyclic codes. 
Constant-weight codes have been used in a number of applications,
 including code-division multiple-access (CDMA) systems for optical fibers, automatic-repeat-request error-control systems, 
parallel asynchronous communications, etc. 
A {\em et al} established \cite{Gy} a general theorem to obtain a binary constant-weight cyclic code from a $p$-ary linear cyclic code, 
where $p$ is a prime, by using a representation of $\GF(p)$  as cyclic shifts of a binary $p$-tuple, and constructions were derived for
four classes of binary constant-weight codes. 
Bitan and Etzion \cite{Bitan} constructed optimal constant weight cyclically permutable codes with weight $w$ 
and minimum Hamming distance $2w-2$.
Xing and Ling \cite{Xing} constructed a class of constant-weight codes by employing the narrow ray class groups of algebraic curves.
Chee and Ling \cite{Chee} introduced a new combinatorial construction for $q$-ary constant-weight 
codes which yields several families of optimal codes and asymptotically optimal codes.
The encyclopedic work on the lower bounds, through explicit constructions, of constant-weight codes was presented in \cite{Brouwer}
whereas the collective upper bounds for constant-weight codes was investigated in \cite{Agrell}.
In this paper, we show that a well-known class of $q$-ary BCH codes excluding the all-zero codeword are constant-weight cyclic codes. 
Moreover, we show this class of codes achieve the Plotkin bound (cf. \cite{Agrell, Berlekamp}).

The third and final purpose of the paper is to characterize the proposed expanded generator matrix and parity check matrix. 
In literature, research has mainly focused on the binary realization of Reed-Solomon codes, 
which has been applied in various practices, e.g., in magnetic recording and optical data storage.
Retter showed \cite{Retter} that the ensemble of generalized Reed-Solomon codes achieve the Gilbert-Varsharmov bound, 
which represents the best-known asymptotic lower bound of the ratio of minimum distance
$d$ to code length $n$ that binary codes of any rate exist (cf. \cite{Berlekamp}). 
In \cite{Retter2}, the orthogonality of binary expansions of Reed-Solomon codes is 
characterized in terms of their spectra and the bases used to expand them.
In \cite{Retter3}, it is shown that the binary weight enumerator of a Reed-Solomon codes over $\GF(2^m)$ 
as well as the gaps of weight distribution generally depend on the choice of basis.
The binary weight enumeration of particular realizations of special Reed-Solomon codes has been studied in \cite{Blake, Kasami, Kolev}.
Vardy and Be'ery \cite{Vardy} showed that high-rate Reed-Solomon codes contain BCH subcodes, 
and subsequently exploited this property to reduce the trellis complexity of bit-level soft-decision maximum-likelihood (Viterbi) decoding. 
Seguin \cite{Seguin} characterized the conditions under which an expanded cyclic code is also cyclic. 
The author extended the subcode concept developed in \cite{Vardy} to more general concept of primary component 
(where a subcode is treated as a trivial component).
Based on the generalized concatenated structure presented in \cite{Seguin}, Sakadibara and Kasahara derived a lower bound 
on the minimum distance of expanded cyclic codes.
Hattori {\em et al} \cite{Hattori} characterized the dimension of subspace subcodes of Reed-Solomon codes under characteristic 2. 
Cui and Pei \cite{Cui} entended the Hattori's work to general $\GF(q^m)$ and to generalized Reed-Solomon codes.
In this paper, we show the precise conditions under which a codeword can be presented by a subbasis. 
Our developments immediately reveals an alternative formula of the dimension 
of subspace subcodes of the Reed-Solomon codes defined in \cite{Hattori}. 
Moreover, the developments handily answer some of open problems listed in the end of \cite{Hattori}, 
including the determination of best subspace and extension to general field $\GF(q^m)$.
With the new insights, we present a lower bound on the minimum distance of an expanded cyclic code, 
exploiting the generalized concatenated structure 
which can be viewed as an improvement over the bound given in \cite{Sak}. 
In \cite{Sak}, the minimum distance of an outer code is shown to be bounded by the largest number
of consecutive conjugacy elements, whereas our developments provide true minimum distance of the outer code, 
which effectively takes into account for the basis realization.
We also show that the binary image of Reed-Solomon codes is asymptotically ``bad", 
in the sense that the ratio of minimum distance over code length diminishes with code length going to infinity.
It overturns the well-known conjecture that they are ``good" codes (cf. \cite{El-Khamy}) and deviates from 
the ensemble of generalized Reed-Solomon codes which asymptotically achieves the Gilbert-Varshamov bound \cite{Retter}.

\section{Description of Expanded Cyclic Codes}

Denote by $\GF(q^m)$  a Galois field, where $q$ is a power of a prime, and $\GF^*(q^m)\eqdef \GF(q^m)\backslash \{0\}$.
 Let $\alpha$ denote a primitive element in $\GF(q^m)$. Let 
\begin{equation}
G(x)=(x-\alpha_1)(x-\alpha_2)(x-\alpha_3)...(x-\alpha_R)  \label{G-poly}
\end{equation}
be the generator polynomial of the primitive cyclic code 
${\mathcal C}(N, K)$, where $N=q^m-1$ and $R=N-K$. 

It is well-known that the parity check matrix can be represented in the form of (cf. \cite{Berlekamp})
\begin{equation}
\bH(\alpha_1, \alpha_2, \ldots, \alpha_R)=\left[\begin{array}{lllllllll}
1 & \alpha_1^1   & \alpha_1^2   &  \alpha_1^3 & \ldots          &   \alpha_1^{N-2}  &   \alpha_1^{N-1}\\
1 & \alpha_2^1   & \alpha_2^2   &  \alpha_2^3 & \ldots          &   \alpha_2^{N-2}  &   \alpha_2^{N-1}\\
1 & \alpha_3^1   & \alpha_3^2   &  \alpha_3^3 & \ldots          &   \alpha_3^{N-2}  &   \alpha_3^{N-1}\\
\vdots &\vdots & \vdots     & \vdots      & \ddots              & \vdots            & \vdots          \\
1 & \alpha_R^1   & \alpha_R^2   &  \alpha_R^3 & \ldots          &   \alpha_R^{N-2}  &   \alpha_R^{N-1}\\
\end{array} \right].  \label{symbol-parity-matrix}
\end{equation}

Denote
\begin{equation}
\bg(\gamma) \eqdef [1,\; \gamma, \; \gamma^2, \; \ldots,\; \gamma^{N-1}],
\end{equation}
and its corresponding polynomial
\begin{equation}
g_{\gamma}(x)\eqdef 1+\gamma x +\gamma^{2} x^2+ \ldots+\gamma^{N-1} x^{N-1}.
\end{equation}
It can be easily shown that 
\begin{eqnarray}
g_{\gamma}(x)&=& \frac{(\gamma x)^N -1}{\gamma x-1} \nonumber \\
      &=& \frac{(\gamma x -1)(\gamma x-\alpha)(\gamma x -\alpha^2) \ldots (\gamma x -\alpha^{N-1})}{\gamma x-1}\nonumber\\
      &=& (\gamma x-\alpha)(\gamma x -\alpha^2) \ldots (\gamma x -\alpha^{N-1}) \nonumber\\
      &=& \gamma^{(N-1)}(x-\gamma^{-1} \alpha^{1}) (x-\gamma^{-1}\alpha^{2}) \ldots (x-\gamma^{-1}\alpha^{N-1}) \nonumber\\
      &=& \gamma^{-1}(x-\gamma^{-1} \alpha^{1}) (x-\gamma^{-1}\alpha^{2}) \ldots (x-\gamma^{-1}\alpha^{N-1}) \label{g-gamma}
 \end{eqnarray}
where the second ``=" is due to
$$x^N-1=(x-\alpha^0)(x-\alpha^1)(x-\alpha^2)\ldots (x-\alpha^{N-2}) (x-\alpha^{N-1}). $$

\begin{lemma} \label{LEM-G-poly-matrix}
Let 
\begin{equation}
\{\gamma_i: \; 1\leq i\leq K\} \eqdef \GF^*(q^m)\backslash \{\alpha_j^{-1}: 1\leq j\leq R\} 
\end{equation}
Then, the ${\mathcal C}(N, K)$ code defined by \eqref{G-poly} has the following generator matrix
\begin{equation}
\bG(\gamma_1, \gamma_2, \ldots, \gamma_K) =\left[\begin{array}{llllllll}
1 & \gamma_1^{1} & \gamma_1^{2} & \gamma_1^{3} & \ldots & \gamma_1^{N-2} & \gamma_1^{N-1} \\
1 & \gamma_2^{1} & \gamma_2^{2} & \gamma_2^{3} & \ldots & \gamma_2^{N-2} & \gamma_2^{N-1} \\
1 & \gamma_3^{1} & \gamma_3^{2} & \gamma_3^{3} & \ldots & \gamma_3^{N-2} & \gamma_3^{N-1} \\
\vdots &\vdots & \vdots     &\vdots      & \ddots & \vdots           &\vdots     \\
1 & \gamma_K^{1} & \gamma_K^{2} & \gamma_K^{3} & \ldots & \gamma_K^{N-2} & \gamma_K^{N-1} 
\end{array} \right].  \label{symbol-G-matrix}
\end{equation}
\end{lemma}
{\em Proof:} Evidently, the above matrix exhibits full rank due to the Vandermonde property. 
On the other hand, as indicated by \eqref{g-gamma}, the polynomials associated with each row of $\bG$
contain roots $\alpha_1$, $\alpha_2$, $\alpha_3$, \ldots, $\alpha_R$, subsequently
divide $G(x)$. \hfill $\Box\Box$

It is worth clarifying that a generator polynomial is directly associated with a parity check matrix, 
whereas a parity check polynomial is directly associated with a generator matrix.

\begin{corollary}
The Reed-Solomon code defined by the generator polynomial
\begin{equation}
G(x)=(x-\alpha^{\delta})  (x-\alpha^{\delta+1}) (x-\alpha^{\delta+2}) \ldots (x-\alpha^{\delta+R-1})
\end{equation}
has generator matrix $\bG(\alpha^{-\delta+1}, \alpha^{-\delta+2}, \ldots, \alpha^{-\delta+K})$,
 as defined in \eqref{symbol-G-matrix}.
\end{corollary}

Let $\{\beta_1, \beta_2, \ldots, \beta_m\}$ be a basis of $\GF(q^m)$. 
An element $\gamma\in \GF(q^m)$ can be decomposed in form of
\begin{equation}
\gamma=\mu_1\beta_1+\mu_2\beta_2+\ldots+\mu_m\beta_m,  \label{decomp-gamma}
\end{equation}
where $\mu_i\in \GF(q)$.

The following theorem presents an explicit construction of generator matrix and parity check matrix of an expanded code,
which maintains the symbol-wise algebraic structure and thus keeps many important original properties.

\begin{theorem} 
$(i)$. Let $\mC(N, K)$ be defined in  $\GF(q^m)$ and with generator matrix
$$\bG=[\bg_1, \; \bg_2, \; \ldots, \; \bg_K]^T.$$ 
Then, its  expansion code in $\GF(q)$ under a basis $\{\beta_i\}_{i=1}^m$ has the following generator matrix
$$\bG_e=[\beta_1 \bg_1, \; \ldots,\;   \beta_m\bg_1,\; 
 \beta_1 \bg_2, \; \ldots,\;  \beta_m\bg_2, \; \ldots \ldots, \;
\beta_1 \bg_K, \;  \ldots,\;   \beta_m\bg_K]^T.$$
$(ii)$. Let $\mC(N, K)$ be defined in  $\GF(q^m)$ and with parity check matrix
$$\bH^T=[\bh_1, \; \bh_2,\; \bh_3,\; \ldots,\; \bh_{N}]^T.$$
Then, its  expansion code in $\GF(q)$ under a basis  $\{\beta_i\}_{i=1}^m$ has the following parity check matrix
$$\bH^T_e=[\beta_1 \bh_1,\; \ldots,\; \beta_m\bh_1,\; 
\beta_1 \bh_2,    \;\ldots,\;   \beta_m\bh_2,\;            \ldots \ldots,\; 
\beta_1 \bh_{N},\; \ldots,\;   \beta_m\bh_{N}]^T.$$
\end{theorem}
Note that the subscript ``$_e$" is used to denote the corresponding expansion. \\
{\em Proof: } $(i)$. We first show the matrix $\bG_e$ is full rank through contradiction. Let 
$$\sum_{j=1}^m \bg_1\nu_{1, j}\beta_j + \sum_{j=1}^m \bg_2\nu_{2, j} \beta_j + \ldots
+ \sum_{j=1}^m \bg_K\nu_{K, j} \beta_j=0,$$
where $\nu_{i, j}\in \GF(q)$ and at least one $\nu_{i, j}$ is nontrivial, and $\bg_i\beta_j$ is viewed as 
an $mN$-dimension vector over the base field $\GF(q)$. 
Let 
$$\theta_i = \sum_{j=1}^m \nu_{i, j}\beta_j, \hspace{0.25in} i=1, 2, \ldots, K.$$
Then, we have
$$\theta_1\bg_1+ \theta_2\bg_2 +\ldots+ \theta_K\bg_K=0,$$
which is contradictory to Lemma~1 where rows in \eqref{symbol-G-matrix} are linearly independent
(herein $\theta_i\bg(\gamma_i)$ is viewed as an $N$-dimension vector over the extension field $\GF(q^m)$).\\
On the other hand, a  codeword can be represented by the linear combination of rows in \eqref{symbol-G-matrix}, 
say, 
$$\bc=\theta'_1\bg_1+ \theta'_2\bg_2 +\ldots+ \theta'_K\bg_K.$$
Note $\theta'_i$ can be represented by the basis $\beta_j$, $j=1, 2, \ldots, m,$ say,
$$\theta'_i=\sum_{j=1}^m \nu'_{i, j}\beta_j, \hspace{0.25in} i=1, 2, \ldots, K.$$
Then, the expansion of the codeword $\bc$ can be decomposed as rows of expanded generator matrix defined in \eqref{binary-G-matrix} 
$$\bc= \sum_{j=1}^m \bg_1\nu'_{1, j}\beta_j + \sum_{j=1}^m \bg_2\nu'_{2, j} \beta_j + \ldots
+ \sum_{j=1}^m \bg_K\nu'_{K, j} \beta_j.$$
The proof of ($ii$) follows the observation below
\begin{eqnarray*}
\aline{} [h_1, \; h_2,\; h_3,\; \ldots,\; h_N] \cdot [c_1,\; c_2, \;c_3,\; \ldots, \; c_N]^T \\
\aline{=}\sum_{i=1}^N h_i c_i \\
\aline{=}\sum_{i=1}^N h_i \sum_{j=1}^m \mu_j(c_i) \beta_j \\
\aline{=}\sum_{i=1}^N \sum_{j=1}^m \mu_j(c_i) h_i\beta_j \\
\aline{=} [h_1\beta_1,\; \ldots, \;h_1\beta_m,\; \ldots,\; h_N\beta_1,\; \ldots,\; h_N\beta_m]
\cdot [\mu_1(c_1),\; \ldots,\; \mu_m(c_1),\; \ldots,\; \mu_1(c_N),\; \ldots,\; \mu_m(c_N)]^T,
\end{eqnarray*}
where $\mu_i(\gamma)\in\GF(q)$ denotes the coefficient associated with $\beta_i$ in the decomposition of $\gamma\in\GF(q^m)$.
\hfill $\Box\Box$ 

\begin{corollary} \label{TH-binary-G-matrix}
Let $\beta_1, \beta_2, \ldots, \beta_m$ be a basis of $\GF(q^m)$. \\
$(i)$. The generator matrix of the expansion of the cyclic code defined by \eqref{symbol-G-matrix} is
\begin{equation}
\bG_e(\gamma_1, \gamma_2, \ldots, \gamma_K)=\left[\begin{array}{ccc ccc ccc}
\beta_1 & \gamma_1^{1}\beta_1 & \gamma_1^{2}\beta_1 & \gamma_1^{3}\beta_1 & \ldots & \gamma_1^{N-2}\beta_1 & \gamma_1^{N-1}\beta_1 \\
\vdots &\vdots & \vdots     &\vdots      & \ddots & \vdots           &\vdots     \\
\beta_m & \gamma_1^{1}\beta_m & \gamma_1^{2}\beta_m & \gamma_1^{3}\beta_m & \ldots & \gamma_1^{N-2}\beta_m & \gamma_1^{N-1}\beta_m \\
\beta_1 & \gamma_2^{1}\beta_1 & \gamma_2^{2}\beta_1 & \gamma_2^{3}\beta_1 & \ldots & \gamma_2^{N-2}\beta_1 & \gamma_2^{N-1}\beta_1 \\
\vdots &\vdots & \vdots     &\vdots      & \ddots & \vdots           &\vdots     \\
\beta_m & \gamma_2^{1}\beta_m & \gamma_2^{2}\beta_m & \gamma_2^{3}\beta_m & \ldots & \gamma_2^{N-2}\beta_m & \gamma_2^{N-1}\beta_m \\
\vdots &\vdots & \vdots     &\vdots      & \ddots & \vdots           &\vdots     \\
\vdots &\vdots & \vdots     &\vdots      & \ddots & \vdots           &\vdots     \\
\beta_1 & \gamma_K^{1}\beta_1 & \gamma_K^{2}\beta_1 & \gamma_K^{3}\beta_1 & \ldots & \gamma_K^{N-2}\beta_1 & \gamma_K^{N-1}\beta_1 \\
\vdots &\vdots & \vdots     &\vdots      & \ddots & \vdots           &\vdots     \\
\beta_m & \gamma_K^{1}\beta_m & \gamma_K^{2}\beta_m & \gamma_K^{3}\beta_m & \ldots & \gamma_K^{N-2}\beta_m & \gamma_K^{N-1}\beta_m \\
\end{array} \right]  \label{binary-G-matrix}
\end{equation} 
$(ii)$. The parity check matrix of the expansion of the cyclic code defined by \eqref{G-poly} is 
\begin{equation}
\bH_e^T(\alpha_1, \alpha_2, \ldots, \alpha_R)=\left[\begin{array}{ccc c ccc}
\beta_1   & \beta_1  & \beta_1 & \ldots  & \beta_1 & \beta_1 \\
\vdots    &\vdots    & \vdots     & \ddots       & \vdots           &\vdots     \\
\beta_m & \beta_m & \beta_m & \ldots & \beta_m & \beta_m \\
\beta_1\alpha_1   & \beta_1\alpha_2  & \beta_1\alpha_3 & \ldots & \beta_1\alpha_{R-1} & \beta_1\alpha_{R} \\
\vdots    &\vdots    & \vdots     & \ddots       & \vdots          &\vdots     \\
\beta_m\alpha_1 & \beta_m\alpha_2 & \beta_m\alpha_3 & \ldots & \beta_m\alpha_{R-1} & \beta_m\alpha_R \\
\vdots    &\vdots    & \vdots     & \ddots       & \vdots           &\vdots     \\
\vdots    &\vdots    & \vdots     & \ddots       & \vdots           &\vdots     \\
\beta_1\alpha_1^{N-1}   & \beta_1\alpha_2^{N-1}  & \beta_1\alpha_3^{N-1} & \ldots & \beta_1\alpha_{R-1}^{N-1} & \beta_1\alpha_{R}^{N-1} \\
\vdots    &\vdots    & \vdots     & \ddots       & \vdots           &\vdots     \\
\beta_m\alpha_1^{N-1} & \beta_m\alpha_2^{N-1} & \beta_m\alpha_3^{N-1} & \ldots & \beta_m\alpha_{R-1}^{N-1} & \beta_m\alpha_R^{N-1} \\
\end{array} \right]  \label{binary-H-matrix}
\end{equation}
\end{corollary}

In Section IV we will characterize the expanded codes based on the proposed expanded generator matrix 
(whereas the analysis straightforwardly applies to the proposed expanded parity check matrix).


\section{A Class of Constant-Weight Cyclic Codes}

The following lemma identifies a subfield element (cf. \cite{Lidl}).
\begin{lemma}  \label{LMsubfield}
An element $\gamma$ in $\GF(q^m)$ lies in the subfield $\GF(q)$ if and only if $\gamma^{q}=\gamma$. 
\end{lemma}

For instance, $\gamma=\alpha^{33} \in \GF(2^{10})$ lies in the subfield $\GF(2^5)$, as $\gamma^{32}=\gamma$.

Let $p_\gamma(x)$ denote the minimal polynomial of $\gamma \in \GF(q^m)$, 
which is defined as the minimum-degree nominal polynomial which has all coefficients pertaining to $\GF(q)$ 
and contains the root $\gamma$. 
Let ${m_\gamma}$ be the minimal dimension of $\gamma$, 
which is defined as the minimum number satisfying $\gamma^{q^{m_\gamma}}=\gamma$ 
(note that $\gamma$ can be represented by an $m_\gamma$-dimensional vector in $\GF(q)$). 
It is worth noting that $m_\gamma$ is a factor of $m$. \\
It is well-known that (cf. \cite{Lidl})
 the minimal polynomial of $\gamma\in \GF(q^m)$ over $\GF(q)$ can be explicitly expressed by
\begin{equation}
p_\gamma(x)=(x-\gamma)(x-\gamma^q)(x-\gamma^{q^2})\ldots(x-\gamma^{q^{{m_\gamma}-1}}).  \label{min-poly}
\end{equation}
where $m_\gamma$ denotes the minimal dimension of $\gamma$. Moreover,
 the conjugacy class, 
\begin{equation}
\phi(\gamma)\eqdef \{ \gamma, \; \gamma^{q},\;  \gamma^{q^2},\; \ldots, \; \gamma^{q^{{m_\gamma}-1}}   \} \label{def-conj}
\end{equation}
share the minimal polynomial $p_\gamma(x)$ (cf. \cite{Lidl}).

Denote by $w$ Hamming weight and $w(\bc)$ the Hamming weight of a vector $\bc$. 
Moreover, denote by $w_\gamma(\bc)$ the weight of $\bc$ contributed by $\gamma$, i.e., $w_\gamma(\bc)\eqdef |\{i: c_i=\gamma\}|$.
The following theorem identifies a class of constant-weight codes.
\begin{theorem} \label{TH-equalWt}
Let $\gamma$ be a non-subfield element in $\GF(q^m)$ and $\mC(N, m)$ be associated with
the generator polynomial $G(x)=\frac{x^{N}-1}{p_{\gamma}(x)}$, 
where $p_\gamma(x)$ is the minimal polynomial of $\gamma$ over $\GF(q)$ as defined in \eqref{min-poly}. 
Then, $\mC^*\eqdef\mC\backslash\{{\boldsymbol 0}\}$ is a code of constant weight $q^{m-1}(q-1)$. 
Moreover, each element of $\GF^*(q)$ appears exactly $q^{m-1}$ times in each codeword.
\end{theorem}
{\em Proof: } 
We observe that the generator polynomial $G(x)$ contains consecutive roots, $\gamma^{q^{m-1}+1}$, 
$\gamma^{q^{m-1}+2}$, \ldots, $\gamma^{N-1}$, $\gamma^{N}=1$. Thus, its minimum distance is at least $q^m-q^{m-1}$. 

On the other hand, note that $G'(x)=\frac{x^{N}-1}{(x-1)p_{\gamma}(x)}$ contains the consecutive roots,
 $\gamma^{q^{m-1}+1}$, $\gamma^{q^{m-1}+2}$, \ldots, $\gamma^{N-1}$. 
Thus, the code $\mC'(N, m+1)$ associated with the generator polynomial $G'(x)$ has minimum distance at least $q^{m}-q^{m-1}-1$. 
Let  ${\bf c}$  be a codeword in $\mC^*$. 
Assume that the nonzero element $\nu^* \in \GF^*(q)$ contributes the most weight to $\bc$, i.e., 
$$w_{\nu^*}(\bc)\geq w_{\nu}(\bc),\hspace{0.2in} \forall\; \nu\in \GF^*(q).$$
Since its Hamming weight is at least $q^m-q^{m-1}$, we have 
$$w_{\nu^*}(\bc)\geq \frac{q^m-q^{m-1}}{q-1}=q^{m-1}.$$ 
We observe that 
$$\bc'=\bc-\nu^*\cdot {\boldsymbol 1} $$ 
is a valid codeword in $\mC'$, where ${\boldsymbol 1}$ denotes the all-one codeword. 
Note $\bc'$ flips all zero elements of $\bc$ to $-\nu^*$ while all $\nu^*$ elements of $\bc$ to zero.
Therefore, we obtain
$$ q^m-q^{m-1}-1 \leq w(\bc')=w(\bc)+w_0(\bc)-w_{\nu^*}(\bc)=q^m-1 -w_{\nu^*}(\bc)$$
which immediate manifests $w_{\nu^*}(\bc)\leq q^{m-1}$. Consequently, it holds $w_{\nu^*}(\bc)=q^{m-1}$. 
Finally, the property $w(\bc)\geq q^{m-1}(q-1)$ holds if and only if 
$$w_\nu(\bc)=q^{m-1}, \hspace{0.3in}  \forall \;  \nu \in \GF^*(q), $$
and subsequently,  
$$w(\bc)=q^{m-1}(q-1), \hspace{0.2in} \forall \bc, $$  
where each of $q-1$ elements in $\GF^*(q)$ equally contributes weight $q^{m-1}$.   
The theorem is concluded.  \hfill $\Box\Box$ 

Let $\alpha$ be a primitive element in $\GF(2^4)$.  Following Theorem~\ref{TH-equalWt}, 
the nonzero codewords associated with the generator polynomial 
$\frac{x^{15}-1}{p_{\alpha^{-1}}(x)}$ have constant weight $2^{4-1}=8$, as listed below.
$$
\begin{array}{ccc}
\left[\;1\; 1\; 1\; 1\; 0\; 1\; 0\; 1\; 1\; 0\; 0\; 1\; 0\; 0\; 0\;  \right] &
\left[\;0\; 1\; 1\; 1\; 1\; 0\; 1\; 0\; 1\; 1\; 0\; 0\; 1\; 0\; 0\;  \right] &
\left[\;1\; 0\; 0\; 0\; 1\; 1\; 1\; 1\; 0\; 1\; 0\; 1\; 1\; 0\; 0\;  \right] \\
\left[\;0\; 0\; 1\; 1\; 1\; 1\; 0\; 1\; 0\; 1\; 1\; 0\; 0\; 1\; 0\;  \right] &
\left[\;1\; 1\; 0\; 0\; 1\; 0\; 0\; 0\; 1\; 1\; 1\; 1\; 0\; 1\; 0\;  \right] &
\left[\;0\; 1\; 0\; 0\; 0\; 1\; 1\; 1\; 1\; 0\; 1\; 0\; 1\; 1\; 0\;  \right] \\
\left[\;1\; 0\; 1\; 1\; 0\; 0\; 1\; 0\; 0\; 0\; 1\; 1\; 1\; 1\; 0\;  \right] &
\left[\;1\; 1\; 1\; 0\; 1\; 0\; 1\; 1\; 0\; 0\; 1\; 0\; 0\; 0\; 1\;  \right] &
\left[\;0\; 0\; 0\; 1\; 1\; 1\; 1\; 0\; 1\; 0\; 1\; 1\; 0\; 0\; 1\;  \right] \\
\left[\;1\; 0\; 0\; 1\; 0\; 0\; 0\; 1\; 1\; 1\; 1\; 0\; 1\; 0\; 1\;  \right] &
\left[\;0\; 1\; 1\; 0\; 0\; 1\; 0\; 0\; 0\; 1\; 1\; 1\; 1\; 0\; 1\;  \right] &
\left[\;1\; 1\; 0\; 1\; 0\; 1\; 1\; 0\; 0\; 1\; 0\; 0\; 0\; 1\; 1\;  \right] \\
\left[\;0\; 0\; 1\; 0\; 0\; 0\; 1\; 1\; 1\; 1\; 0\; 1\; 0\; 1\; 1\;  \right] &
\left[\;1\; 0\; 1\; 0\; 1\; 1\; 0\; 0\; 1\; 0\; 0\; 0\; 1\; 1\; 1\;  \right] &
\left[\;0\; 1\; 0\; 1\; 1\; 0\; 0\; 1\; 0\; 0\; 0\; 1\; 1\; 1\; 1\;  \right]
\end{array}$$

\begin{theorem} \label{TH-subequalWt}
Let $\gamma\in \GF(q^m)$ be a subfield element. 
Let $\mC(N, m_\gamma)$ be defined by the generator polynomial $\frac{x^{N}-1}{p_{\gamma}(x)}$, 
where $p_\gamma(x)$ is the minimal polynomial as defined in \eqref{min-poly}.
Then, $\mC^*\eqdef \mC\backslash\{{\boldsymbol 0}\}$ is a code of constant weight
 $q^{m_\gamma-1}(q-1) \frac{q^m-1}{q^{m_\gamma}-1}$, where $m_\gamma$ denotes the minimal dimension of $\gamma$.
Moreover, each element of $\GF^*(q)$ appears exactly $q^{m_\gamma-1}\frac{q^m-1}{q^{m_\gamma}-1}$ times in each codeword.
\end{theorem}
{\em Proof: }
Lemma~\ref{LEM-G-poly-matrix} in conjunction with Corollary~\ref{TH-binary-G-matrix} indicates that the expanded generator matrix is, 
$$\bG_e=[\beta_1\bg(\gamma^{-1}),\; \ldots, \;\beta_m\bg(\gamma^{-1}),\; 
\beta_1\bg(\gamma^{-q}),\; \ldots, \; \beta_m\bg(\gamma^{-q}),\; \ldots,\;
\beta_1\bg(\gamma^{-q^{m_\gamma-1}}),\; \ldots, \; \beta_m\bg(\gamma^{-q^{m_\gamma-1}})]^T. $$
We observe that each row is periodic with duration $q^{m_\gamma}-1$, and thus contains $\frac{q^m-1}{q^{m_\gamma}-1}$ periods (note that 
$m_\gamma=m_{\gamma^{-1}}$).
This shows that each expanded codeword is also periodic and contains $\frac{q^m-1}{q^{m_\gamma}-1}$ periods.
It can be easily seen that $\gamma$ is a primitive element in the subfield $\GF(q^{m_\gamma})$
and each period of an expanded code is exactly a codeword associated with the generator polynomial
$\frac{x^{(q^{m_\gamma}-1)}-1}{p_\gamma(x)}$ defined in the subfield $\GF(q^{m_\gamma})$.
Thus, each period of a code has constant weight $q^{m_\gamma-1}(q-1)$, following Theorem~\ref{TH-equalWt}. \hfill $\Box\Box$

Let $\alpha$ be a primitive element in $\GF(2^6)$. 
Following Theorem~\ref{TH-subequalWt}, the nonzero codewords associated with the generator polynomial 
$\frac{x^{63}-1}{p_{\alpha^{-9}}(x)}$ 
have constant weight $2^{3-1}\cdot \frac{2^6-1}{2^3-1}=36$, as listed below.
$$\begin{array}{c}
\left[\; 1 0 1 1 1 0 0 \;  1 0 1 1 1 0 0 \;  1 0 1 1 1 0 0\;   
		 1 0 1 1 1 0 0 \;  1 0 1 1 1 0 0 \;  1 0 1 1 1 0 0\;
	     1 0 1 1 1 0 0 \;  1 0 1 1 1 0 0 \;  1 0 1 1 1 0 0\;  \right] \\
\left[\; 1 1 1 0 0 1 0 \;  1 1 1 0 0 1 0 \;  1 1 1 0 0 1 0\;  
		 1 1 1 0 0 1 0 \;  1 1 1 0 0 1 0 \;  1 1 1 0 0 1 0\;  
		 1 1 1 0 0 1 0 \;  1 1 1 0 0 1 0 \;  1 1 1 0 0 1 0\;  \right] \\
\left[\; 0 1 0 1 1 1 0 \;  0 1 0 1 1 1 0 \;  0 1 0 1 1 1 0\;  
		 0 1 0 1 1 1 0 \;  0 1 0 1 1 1 0 \;  0 1 0 1 1 1 0\;  
		 0 1 0 1 1 1 0 \;  0 1 0 1 1 1 0 \;  0 1 0 1 1 1 0\;  \right] \\
\left[\; 0 1 1 1 0 0 1 \;  0 1 1 1 0 0 1 \;  0 1 1 1 0 0 1\;
		 0 1 1 1 0 0 1 \;  0 1 1 1 0 0 1 \;  0 1 1 1 0 0 1\;
		 0 1 1 1 0 0 1 \;  0 1 1 1 0 0 1 \;  0 1 1 1 0 0 1\;  \right] \\
\left[\; 1 1 0 0 1 0 1 \;  1 1 0 0 1 0 1 \;  1 1 0 0 1 0 1\;  
		 1 1 0 0 1 0 1 \;  1 1 0 0 1 0 1 \;  1 1 0 0 1 0 1\;  
	     1 1 0 0 1 0 1 \;  1 1 0 0 1 0 1 \;  1 1 0 0 1 0 1\;  \right] \\ 
\left[\; 1 0 0 1 0 1 1 \;  1 0 0 1 0 1 1 \;  1 0 0 1 0 1 1\;    
		 1 0 0 1 0 1 1 \;  1 0 0 1 0 1 1 \;  1 0 0 1 0 1 1\;    
		 1 0 0 1 0 1 1 \;  1 0 0 1 0 1 1 \;  1 0 0 1 0 1 1\;  \right] \\
\left[\; 0 0 1 0 1 1 1 \;  0 0 1 0 1 1 1 \;  0 0 1 0 1 1 1\;  
		 0 0 1 0 1 1 1 \;  0 0 1 0 1 1 1 \;  0 0 1 0 1 1 1\;  
		 0 0 1 0 1 1 1 \;  0 0 1 0 1 1 1 \;  0 0 1 0 1 1 1\; \right]
\end{array} $$

The Plotkin bound asserts that the minimum distance $\dmin$ of any (linear or nonlinear) code
which has $A$ codewords of length $N$ over the alphabet of size $q$ is bounded by (cf. \cite{Berlekamp})
$$\dmin\leq \frac{N(q-1)}{q-qA^{-1} }\; .$$
\begin{theorem}
$(i)$. The cyclic code $\mC(N, m_\gamma)$ associated with generator polynomial $\frac{x^N-1}{p_\gamma(x)}$ 
exactly matches the Plotkin bound,  where $m_\gamma$ denotes the minimal dimension of $\gamma$.\\
$(ii)$. Given that $\gamma$ is a non-subfield element in $\GF(q^m)$, 
the cyclic code $\mC(N, m+1)$ associated with generator polynomial $\frac{x^N-1}{(x-1)p_\gamma(x)}$ 
matches the Plotkin bound.\\
$(iii)$. Given that $\gamma$ is a non-subfield element in $\GF(q^m)$, 
the constant-weight cyclic code $\mC^*(N, m)$ associated with the generator polynomial $\frac{x^N-1}{p_\gamma(x)}$ 
matches the Plotkin bound. 
\end{theorem}

{\em Proof: } $(i)$. When $\gamma$ is a non-subfield element, we have 
$$\frac{N(q-1)}{q-qA^{-1} } = \frac{(q^m-1)(q-1)}{ q- q(q^m)^{-1} } = q^{m-1}(q-1)=\dmin,$$
where the code size $A=q^m$.
When $\gamma$ is a subfield element, we again have
$$\frac{N(q-1)}{q-qA^{-1} } = \frac{(q^m-1)(q-1)}{ q- q(q^{m_\gamma})^{-1} } 
= q^{m_\gamma-1}(q-1) \frac{q^m-1}{q^{m_\gamma}-1} = \dmin,$$
where the code size $A=q^{m_\gamma}$.

$(ii)$. Note that the minimum distance is precisely $q^m-q^{m-1}-1$.
Consequently,
 $$\left\lfloor \frac{N(q-1)}{q-qA^{-1} } \right\rfloor = \left\lfloor \frac{(q^m-1)(q-1)}{ q- q(q^{m+1})^{-1} }  \right\rfloor 
= q^{m-1}(q-1)-1 +  \left\lfloor  \frac{2-q^{-1}-q^{-m}}{q-q^{-m}} \right\rfloor = q^{m-1}(q-1)-1=\dmin, $$
where the code size $A=q^{m+1}$. 

The proof of part $(ii)$ follows
$$\left\lfloor \frac{N(q-1)}{q-qA^{-1} } \right\rfloor= \left\lfloor \frac{(q^m-1)(q-1)}{ q- q(q^m-1)^{-1} }  \right\rfloor
= \left\lfloor q^{m-1}(q-1) + \frac{q-1}{q(q^m-2)} \right\rfloor = q^{m-1}(q-1)=\dmin,$$
where the code size $A=q^m-1$.  \hfill $\Box\Box$


\section{Characterization of Expanded Cyclic Codes}

In this section, we  carry out analysis on expanded generator matrix $\bG_e$.
It is straightforward to show that all results hold in analogue to expanded parity check matrix $\bH_e$. 

Given a basis $\{\beta_i\}_{i=1}^m$ in $\GF(q^m)$, denote
by $\mu_i(\gamma)\in \GF(q) $ the function of $\gamma$ that represents its expansion associated with basis $\beta_i$,
i.e., the value $\mu_i$ in \eqref{decomp-gamma}.
For brevity, the function $\mu_i(\cdot)$ is also applied to a vector ${\bf y}=[y_1,\; y_2, \; \ldots,\; y_n]\in \GF(q^m)^n$, such that
\begin{equation}
\mu_i({\bf y}) \eqdef [\mu_i(y_1), \;\mu_i(y_2), \;\mu_i(y_3),\; \ldots, \; \mu_i(y_n)].
\end{equation}
and to a polynomial $f(x)=f_0 + f_1 x + f_2 x^2+\ldots+ f_n x^n \in \GF(q^m)[x]$, such that
\begin{equation}
\mu_i( f(x) ) \eqdef \mu_i(f_0) + \mu_i(f_1) \cdot x + \mu_i(f_2)\cdot  x^2 + \ldots + \mu_i(f_n) \cdot x^n.
\end{equation}
Let 
\begin{equation}
\beta_{i_1}\beta_{i_2}...\beta_{i_k}=\sum_{j=1}^m \mu^{(i_1, i_2, \ldots, i_k)}_j\cdot \beta_j, \label{beta-ij}
\end{equation}
where $\mu^{(i_1, i_2, \ldots, i_k)}_j$, $j=1, 2, \ldots, m$, are regarded as constants since $\{ \beta_i \}_{i=1}^m$ are treated as known parameters, 
and for brevity
\begin{equation}
\beta_i^{q^s}=\sum_{j=1}^m \mu_j^{(i[q^s])} \beta_j.
\end{equation}

\begin{lemma}  \label{non-subfield}
Given a non-subfield element $\gamma$ in $\GF^*(q^m)$, 
the elements of $\bg(\gamma)$ cannot be generated by a subbasis. 
\end{lemma}
Note that subbasis is a weaker concept than subfield. The basis of a subfield can be expanded to represent the whole field,
thus a subfield corresponds a subbasis, whereas the a subbasis is not necessarily associated with a subfield.

{\em Proof: } We prove it by contradiction. Let $\{\beta_1, \beta_2, \ldots, \beta_m\}$ be the basis of $\GF(q^m)$ 
and $\{\beta_{i_1}, \beta_{i_2}, \ldots, \beta_{i_k}\}$ ($k<m$) be the subbasis for the elements of $\bg(\gamma)$.
Consequently, the linear span (under addition operation) of the multiplicative group 
\{1, $\gamma$, $\gamma^2$, \ldots, $\gamma^{N-1}$\} is a field, with 
basis $\{\beta_{i_1}$, $\beta_{i_2}$, \ldots, $\beta_{i_k}\}$. 
It means that the span (under addition) of $\{\beta_{i_1}$, $\beta_{i_2}$, \ldots, $\beta_{i_k}\}$, which has $q^k$ elements, is a proper subfield of $\GF(q^m)$. 
It follows that $\gamma$ is subfield element, violating the assumption.     \hfill $\Box\Box$

According to \eqref{g-gamma}, $\beta_i g_{\gamma}(x)$ can be divided into 
$$\beta_i g_{\gamma}(x)= \frac{x^{N}-1}{p_{\gamma^{-1}}(x)} \beta_i \chi_{\gamma}(x),$$
where $\frac{x^{N}-1}{p_{\gamma^{-1}}(x)} \in \GF(q)[x]$, and 
\begin{equation}
\chi_{\gamma}(x)=\gamma^{-1} (x-\gamma^{-q})(x-\gamma^{-q^2})\ldots(x-\gamma^{-q^{-1+m_\gamma}})  \label{def-chi}
\end{equation}
where ${m_\gamma}=m_{\gamma^{-1}}$ (recall that it is defined as the smallest number such that $\gamma^{q^{m_\gamma}}=\gamma$).
Decompose $\beta_i g_{\gamma}(x)$ into
$$\beta_i g_{\gamma}(x)= \sum_{j=1}^m \beta_j\cdot \mu_j(\beta_i\chi_{\gamma}(x))\frac{x^{N}-1}{p_{\gamma^{-1}}(x)}.$$
The above expression immediately manifests that 
\begin{equation}
\mu_j(\beta_i g_\gamma(x))=\mu_j(\beta_i\chi_{\gamma}(x))\frac{x^{N}-1}{p_{\gamma^{-1}}(x)}, \label{component-g-x}
\end{equation}
The following theorem summarizes the significant property of $\bG_e(\gamma)$.
\begin{theorem}   \label{TH-single-term}
$(i)$. Given $\gamma$ a non-subfield element in $\GF^*(q^m)$, no nontrivial codeword associated with the generator polynomial $\bG_e(\gamma)$ 
 can be represented by a subbasis.
Moreover, the component words $\mu_j(\beta_i\bg(\gamma))$, $1\leq i, j\leq m$, 
are codewords associated with the generator polynomial $\frac{x^{N}-1}{p_{\gamma^{-1}}(x)}$ 
and  exhibit constant Hamming weight $q^{m-1}(q-1)$. \\
$(ii)$. Given $\gamma$ a subfield element with minimal dimension $m_\gamma<m$,
the component words $\mu_j(\beta_i\bg(\gamma))$, $1\leq i, j\leq m$, 
are codewords associated with the generator polynomial $\frac{x^{N}-1}{p_{\gamma^{-1}}(x)}$,
and  exhibit Hamming weight $q^{m_\gamma-1}(q-1) \frac{q^m-1}{q^{m_\gamma}-1}$ or zero. 
Moreover, a component codeword $\mu_j(\beta_i\bg(\gamma))$ is the all-zero word if and only if $j\ne i$ and 
$\beta_i\GF(q^{m_\gamma})$ can be represented by the subbasis 
$\{\beta_1, \ldots, \beta_{j-1}, \beta_{j+1}, \ldots, \beta_{m}\}$.
\end{theorem}
{\em Proof: } We shall only show the part related to subbasis representation. \\
$(i)$. A codeword $\bc$ can be expressed as $\bc=\theta\bg(\gamma)$ for some $\theta\in \GF^*(q^m)$.
Assume that $\theta\bg(\gamma)$ is generated by a subbasis $\{\beta_{i_1}$, $\beta_{i_2}$, \ldots, $\beta_{i_k}\}$ out of $\{\beta_i\}_{i=1}^m$ ($k<m$), 
then, $\bg(\gamma)$ is generated by the subbasis $\{\theta^{-1}\beta_{i_1}$, $\theta^{-1}\beta_{i_2}$, \ldots, $\theta^{-1}\beta_{i_k} \}$ 
(out of the alternative basis $\{\theta^{-1}\beta_i\}_{i=1}^m$).
This clearly conflicts to Lemma~\ref{non-subfield}, which asserts that $\bg(\gamma)$ cannot be generated by a subbasis. \\
$(ii)$. Clearly, a straightforward equivalence is that a component codeword $\mu_j(\beta_i\bg(\gamma))$ is the all-zero word 
if and only if $\beta_i\bg(\gamma)$ can be represented by the subbasis 
$\{\beta_1, \ldots, \beta_{j-1}, \beta_{j+1}, \ldots, \beta_{m}\}$.
The condition $j\ne i$ is due to the fact that $\beta_j$ can not be represented by
the subbasis $\{\beta_1, \ldots, \beta_{j-1}, \beta_{j+1}, \ldots, \beta_{m}\}$.
On the other hand, Lemma~\ref{non-subfield} indicates that the field $\GF(q^{m_\gamma})$ is the closure of $\bg(\gamma)$ 
under the addition operation, therefore, if $\beta_i\bg(\gamma)$ is represented by a subbasis, 
then $\beta_i\GF(q^{m_\gamma})$ is also represented by the subbasis. \hfill $\Box\Box$

Note that the constant-weight characterization follows Theorem~\ref{TH-equalWt} and $m_\gamma=m_{\gamma^{-1}}$. 
We give two examples to illustrate part $(ii)$. 
Let the composite basis $\{\beta_1$, $\beta_2$, $\beta_3$, $\beta_4\}=\{1$, $\alpha^5$, $\alpha$, $\alpha^6\}$ 
be employed to expand $\bg(\alpha^{10})$ in $\GF(2^4)$. We have 
$$\mu_i(\beta_j\bg(\alpha^{10}))={\boldsymbol 0}, \hspace{0.2in} i=3, 4, \; j=1, 2, \text{ or } i=1, 2, \; j=3, 4. $$
Alternatively, let the basis be $\{\beta_1$, $\beta_2$, $\beta_3$, $\beta_4\}=\{1+\alpha$, $\alpha^5$, $\alpha$, $\alpha^6\}$,
where the subfield $\GF(2^2)$ is represented by the subbasis $\{1+\alpha$, $\alpha^5$, $\alpha\}$. Consequently, we have,
$$\mu_i(\beta_j\bg(\alpha^{10}))={\boldsymbol 0}, \hspace{0.2in} i=4, \; j=2, \text{ or } i=1, 2, \; j=3, 4. $$

Note that the above theorem justifies that the density of binary expanded parity check matrix of a Reed-Solomon code is near one half,
due to dominant non-subfield elements whose corresponding density is precisely $\frac{2^{(m-1)}}{2^m-1}$.

\begin{corollary}  \label{CO-common-poly}
If a polynomial $p(x)\in \GF(q)[x]$ divides the generator polynomial $G(x)\in\GF(q^m)[x]$, 
then it also divides the all component word polynomials $\mu_i(c(x))$, $i=1, 2, \ldots, m$,
where $c(x)$ denotes a codeword generated by $G(x)$. Moreover,  
let $d$ be the minimum distance of the code associated with the generator polynomial $p(x)$,
then the weight of a nonzero component codeword, $\mu_i(\bc)$, is at least $d$.
\end{corollary}
{\em Proof: } Let $G(x)=p(x)G'(x)$ and $c(x)=a(x)G(x)$.
In analogue to \eqref{component-g-x}, we have
$$\mu_i(c(x))=\mu_i(a(x)G(x))=\mu_i(a(x)G'(x)p(x))=\mu_i(a(x)G'(x))\cdot p(x)$$
for $i=1, 2, \ldots, m$. \hfill $\Box\Box$

We observe that 
\begin{equation}
\gamma\beta_i= \sum_{j=1}^m \mu_j(\gamma) \beta_j\beta_i, \;\; 1\leq i\leq m.
\end{equation}
Substituting \eqref{beta-ij} into the above expression, we obtain
\begin{eqnarray}
\gamma\beta_i&=&\sum_{l=1}^m \beta_l\sum_{j=1}^m \mu^{(i, j)}_l \mu_j(\gamma) \nonumber \\
&=& \sum_{l=1}^m \beta_l \cdot f_{i, l}(\mu_1(\gamma), \mu_2(\gamma), \ldots, \mu_m(\gamma))
\end{eqnarray}
where the coefficient function
\begin{equation}
f_{i, l}(\mu_1, \mu_2, \ldots, \mu_m)\eqdef \sum_{j=1}^m \mu^{(i, j)}_l \mu_j, \;\; i=1, 2, \ldots, m.
\end{equation}
The following lemma characterizes the properties of linear function $f_{i, j}(\mu_1, \mu_2, \ldots, \mu_m)$.
\begin{lemma}   \label{LMoneElement}
$(i)$. For any given $j$, there doesnot exist nontrivial $\{\nu_i\}_{i=1}^m\in \GF(q)$ such that
  $$\sum_{i=1}^m \nu_i f_{i, j}(\mu_1(\gamma), \mu_2(\gamma), \ldots, \mu_m(\gamma)) =0,\;\;\; \forall \gamma\in\GF^*(q^m).$$
$(ii)$. Given $\gamma$ a non-subfield element in $\GF^*(q^m)$, the $m$ vectors 
$\left[f_{i, j}(\mu_1(1)\right.$, \ldots, $\mu_m(1))$, $f_{i, j}(\mu_1(\gamma)$, \ldots, $\mu_m(\gamma))$,  
\ldots, $f_{i, j}(\mu_1(\gamma^{N-1})$, \ldots, $\left.\mu_m(\gamma^{N-1}))\right]$,
$i=1, 2, \ldots, m$ are linearly independent.  
\end{lemma}
{\em Proof: }   $(i)$. Suppose it is not true, say,
$$\sum_{i=1}^m \nu_i f_{i, j}(\mu_1(\gamma), \mu_2(\gamma), \ldots, \mu_m(\gamma) )=0$$
for arbitrary $\gamma\in\GF^*(q^m)$. Note that
$$\sum_{i=1}^m \nu_i f_{i, j}(\mu_1(\gamma), \mu_2(\gamma), \ldots, \mu_m(\gamma) )
=\sum_{i=1}^m \nu_i \mu_j(\beta_i\gamma)=\mu_j\left(\sum_{i=1}^m \nu_i \beta_i\gamma\right),$$
where the last equality is due to the linearity of $\mu_j$ in $\GF(q)$. 
Since $\gamma$ ranges over $\GF^*(q^m)$, $\gamma\cdot \sum_{i=1}^m \nu_i \beta_i$ ranges over $\GF^*(q^m)$ as well.
However, it is obviously wrong as it indicates that all elements in $\GF(q^m)$ are can be represented by the subbasis 
$\{\beta_1$,\ldots, $\beta_{j-1}$, $\beta_{j+1}$, \ldots, $\beta_m\}$. 

$(ii)$. Suppose it is not true, say, 
$$\sum_{i=1}^m \nu_i \left[f_{i, j}(\mu_1(1), \ldots, \mu_m(1)), 
f_{i, j}(\mu_1(\gamma), \ldots, \mu_m(\gamma)), \ldots,
f_{i, j}(\mu_1(\gamma^{N-1}), \ldots, \mu_m(\gamma^{N-1}))\right] 
={\boldsymbol 0},$$
where $\nu_i\in\GF(q)$, for some $\gamma \in \GF(q^m)$ and $j$. 
This is equivalent to that all elements $\sum_{i=1}^m \nu_i\beta_i \cdot\bg(\gamma)$
 do not contain the basis component $\beta_j$. Lemma~\ref{non-subfield} shows that  
$\sum_{i=1}^m \nu_i\beta_i \cdot \bg(\gamma)$ pertains to a subfield. 
This indicates that 
$$\gamma=\left(\sum_{i=1}^m \nu_i\beta_i \right)^{-1} \cdot \left(\gamma \sum_{i=1}^m\nu_i \beta_i \right),$$ 
where the two terms on the right side are the first and second elements of $\sum_{i=1}^m \nu_i \beta_i \cdot \bg(\gamma)$, respectively. 
This indicates that $\gamma$ is a subfield element, which violates the assumption.   \hfill $\Box\Box$

Note that Lemma~\ref{LMoneElement}.$(ii)$ may not hold true when $\gamma$ belongs to a subfield of $\GF(q^m)$. 
E.g., let $\gamma=\alpha^{5}$ be in the field $\GF(2^4)$, then we have
$$\bg(\alpha^5)=[1, \alpha^{5}, \alpha^{10}, 1, \alpha^{5}, \alpha^{10}, 1, \alpha^{5}, \alpha^{10},   
1, \alpha^{5}, \alpha^{10}, 1, \alpha^{5}, \alpha^{10}]$$
lying in the subfield $\GF(2^2)$.

We observe that
\begin{equation}
\gamma^{q^s} = \sum_{i=1}^m \mu_i^{q^{s}}(\gamma)\cdot \beta^{q^s}_i
=\sum_{i=1}^m \mu_i(\gamma)\cdot \beta^{q^s}_i 
= \sum_{i=1}^m \beta_i  \sum_{j=1}^m \mu^{(j[q^s])}_i \cdot \mu_j(\gamma) .  \label{gamma-power}
\end{equation}

Combining \eqref{gamma-power} and \eqref{beta-ij}, we obtain
\begin{equation}
\gamma^{q^s} \beta_i=\sum_{j=1}^m\beta_j \sum_{k=1}^m \mu^{(i, k)}_j \sum_{l=1}^m \mu^{(l[q^s])}_k \cdot \mu_l(\gamma), 
\end{equation}
which immediately yields
\begin{equation}
\mu_j(\gamma^{q^s} \beta_i)=\sum_{k=1}^m \mu^{(i, k)}_j \sum_{l=1}^m \mu^{(l[q^s])}_k \cdot \mu_l(\gamma). \label{gamma-power-decomp}
\end{equation}
Letting $\gamma=\beta_r$, the above equality becomes
\begin{equation}
\mu_j(\beta_r^{q^s} \beta_i)=\sum_{k=1}^m \mu^{(i, k)}_j \sum_{l=1}^m \mu^{(l[q^s])}_k \cdot \mu_l(\beta_r)= 
\sum_{k=1}^m \mu^{(i, k)}_j \mu^{(r[q^s])}_k,
\end{equation}
where by definition $\mu_l(\beta_r)= 1$ if $l=r$ or 0  otherwise. 
Consequently, \eqref{gamma-power-decomp} can be re-written as
\begin{equation}
\mu_j(\gamma^{q^s} \beta_i)=\sum_{l=1}^m \mu_j( \beta_i\beta_l^{q^s} ) \cdot \mu_l(\gamma). 
\end{equation}
It follows that
\begin{equation}
\mu_j(\beta_i \bg(\gamma^{q^s})) = \mu_j( \beta_i\beta_1^{q^s} ) \cdot \mu_1(\bg(\gamma)) 
+\mu_j( \beta_i\beta_2^{q^s} ) \cdot \mu_2(\bg(\gamma)) +\ldots
+ \mu_j( \beta_i\beta_m^{q^s} ) \cdot \mu_m(\bg(\gamma)). \label{vec-decomp}
\end{equation}

The following two theorems characterize the intrinsic connection between subbasis and conjugate elements.
\begin{theorem}  \label{THconjugacy}
Given  an expanded generator matrix 
$\bG_e( \gamma, \gamma^{q^{s_1}}$, $\gamma^{q^{s_2}}$,  \ldots,   $\gamma^{q^{s_{k-1}}} )$, \\
$(i)$. When $\gamma$ is a non-subfield element in $\GF^*(q^m)$, 
the dimension of the subspace subcode with respect to a subbasis $\{\beta_i\}_{i=1}^m\backslash \{\beta_{i_1}, \beta_{i_2}, \ldots, \beta_{i_t}\}$ is 
\begin{equation}
mk- \mathcal{R}({\boldsymbol \Gamma}),
\end{equation}
where $\mathcal{R}({\boldsymbol \Gamma})$ denotes the maximum number of linearly independent rows of the matrix ${\boldsymbol \Gamma}$ defined as 
\begin{equation}
{\boldsymbol \Gamma}\eqdef \left[
\begin{array}{ccc c ccc}
\mu_{i_1}(\beta_1\beta_1) &  \ldots & \mu_{i_1}(\beta_1\beta_m ) & \ldots \ldots &  
\mu_{i_t}(\beta_1\beta_1 ) &  \ldots & \mu_{i_t}(\beta_1\beta_m )  \\
\vdots & \ddots  & \vdots & \ddots  & \vdots & \ddots & \vdots \\
\mu_{i_1}(\beta_m\beta_1 ) &  \ldots & \mu_{i_1}(\beta_m\beta_m ) & \ldots\ldots &  
\mu_{i_t}(\beta_m\beta_1 ) &  \ldots & \mu_{i_t}(\beta_m\beta_m )  \\
\mu_{i_1}(\beta_1\beta_1^{q^{s_1}}) &  \ldots & \mu_{i_1}(\beta_1\beta_m^{q^{s_1}}) & \ldots \ldots &  
\mu_{i_t}(\beta_1\beta_1^{q^{s_1}}) &  \ldots & \mu_{i_t}(\beta_1\beta_m^{q^{s_1}})  \\
\vdots & \ddots  & \vdots & \ddots  & \vdots & \ddots & \vdots \\
\mu_{i_1}(\beta_m\beta_1^{q^{s_1}}) &  \ldots & \mu_{i_1}(\beta_m\beta_m^{q^{s_1}}) & \ldots\ldots &  
\mu_{i_t}(\beta_m\beta_1^{q^{s_1}}) &  \ldots & \mu_{i_t}(\beta_m\beta_m^{q^{s_1}})  \\
\vdots & \ddots  & \vdots & \ddots  & \vdots & \ddots & \vdots \\
\mu_{i_1}(\beta_1\beta_1^{q^{s_{k-1}}}) &  \ldots & \mu_{i_1}(\beta_1\beta_m^{q^{s_{k-1}}}) &\ldots \ldots &  
\mu_{i_t}(\beta_1\beta_1^{q^{s_{k-1}}}) &  \ldots & \mu_{i_t}(\beta_1\beta_m^{q^{s_{k-1}}})  \\
\vdots & \ddots  & \vdots & \ddots  & \vdots & \ddots & \vdots \\
\mu_{i_1}(\beta_m\beta_1^{q^{s_{k-1}}}) &  \ldots & \mu_{i_1}(\beta_m\beta_m^{q^{s_{k-1}}}) & \ldots\ldots &  
\mu_{i_t}(\beta_m\beta_1^{q^{s_{k-1}}}) &  \ldots & \mu_{i_t}(\beta_m\beta_m^{q^{s_{k-1}}})  \\
\end{array} \right].
\end{equation}
The dimension exhibits the lower bound $m(k-t)$.\\
$(ii)$. When $\gamma$ is a proper subfield element in $\GF^*(q^m)$, such that $\GF(q^{m_\gamma})$ can be represented by the minimal subbasis
$\{\beta_{j_1}, \beta_{j_2}, \ldots, \beta_{j_r}\}$ (where "minimial" means that any of its proper subset fails), 
the dimension of the subspace subcode with respect to a subbasis $\{\beta_i\}_{i=1}^m\backslash \{\beta_{i_1}, \beta_{i_2}, \ldots, \beta_{i_t}\}$ is 
\begin{equation}
mk- \mathcal{R}({\boldsymbol \Gamma}'),
\end{equation}
where $\mathcal{R}({\boldsymbol \Gamma}')$ denotes the maximum number of linearly independent rows of the matrix ${\boldsymbol \Gamma}'$ defined as 
\begin{equation}
{\boldsymbol \Gamma}'\eqdef \left[
\begin{array}{ccc c ccc}
\mu_{i_1}(\beta_1\beta_{j_1}) &  \ldots & \mu_{i_1}(\beta_1\beta_{j_r} ) & \ldots \ldots &  
\mu_{i_t}(\beta_1\beta_{j_1} ) &  \ldots & \mu_{i_t}(\beta_1\beta_{j_r} )  \\
\vdots & \ddots  & \vdots & \ddots  & \vdots & \ddots & \vdots \\
\mu_{i_1}(\beta_m\beta_{j_1} ) &  \ldots & \mu_{i_1}(\beta_m\beta_{j_r} ) & \ldots\ldots &  
\mu_{i_t}(\beta_m\beta_{j_1} ) &  \ldots & \mu_{i_t}(\beta_m\beta_{j_r} )  \\
\mu_{i_1}(\beta_1\beta_{j_1}^{q^{s_1}}) &  \ldots & \mu_{i_1}(\beta_1\beta_{j_r}^{q^{s_1}}) & \ldots \ldots &  
\mu_{i_t}(\beta_1\beta_{j_1}^{q^{s_1}}) &  \ldots & \mu_{i_t}(\beta_1\beta_{j_r}^{q^{s_1}})  \\
\vdots & \ddots  & \vdots & \ddots  & \vdots & \ddots & \vdots \\
\mu_{i_1}(\beta_m\beta_{j_1}^{q^{s_1}}) &  \ldots & \mu_{i_1}(\beta_m\beta_{j_r}^{q^{s_1}}) & \ldots\ldots &  
\mu_{i_t}(\beta_m\beta_{j_1}^{q^{s_1}}) &  \ldots & \mu_{i_t}(\beta_m\beta_{j_r}^{q^{s_1}})  \\
\vdots & \ddots  & \vdots & \ddots  & \vdots & \ddots & \vdots \\
\mu_{i_1}(\beta_1\beta_{j_1}^{q^{s_{k-1}}}) &  \ldots & \mu_{i_1}(\beta_1\beta_{j_r}^{q^{s_{k-1}}}) &\ldots \ldots &  
\mu_{i_t}(\beta_1\beta_{j_1}^{q^{s_{k-1}}}) &  \ldots & \mu_{i_t}(\beta_1\beta_{j_r}^{q^{s_{k-1}}})  \\
\vdots & \ddots  & \vdots & \ddots  & \vdots & \ddots & \vdots \\
\mu_{i_1}(\beta_m\beta_{j_1}^{q^{s_{k-1}}}) &  \ldots & \mu_{i_1}(\beta_m\beta_{j_r}^{q^{s_{k-1}}}) & \ldots\ldots &  
\mu_{i_t}(\beta_m\beta_{j_1}^{q^{s_{k-1}}}) &  \ldots & \mu_{i_t}(\beta_m\beta_{j_r}^{q^{s_{k-1}}})  \\
\end{array} \right].
\end{equation}
\end{theorem}
{\em Proof:} $(i)$. 
\eqref{vec-decomp} indicates that
\begin{equation}
\left[\begin{array} {c}
\mu_{i_1}(\beta_\rho\bg(\gamma^{q^{s_l}})) \\
 \mu_{i_2}(\beta_\rho \bg(\gamma^{q^{s_l}})) \\
\vdots \\
 \mu_{i_t}(\beta_\rho \bg(\gamma^{q^{s_l}}))
\end{array}\right]
=\left[\begin{array}{c c c c}
\mu_{i_1}(\beta_\rho\beta_1^{s_l})  & \mu_{i_1}(\beta_\rho\beta_2^{s_l})  & \ldots & \mu_{i_1}(\beta_\rho\beta_m^{s_l})  \\
\mu_{i_2}(\beta_\rho\beta_1^{s_l})  & \mu_{i_2}(\beta_\rho\beta_2^{s_l})  & \ldots & \mu_{i_2}(\beta_\rho\beta_m^{s_l})  \\
\vdots & \vdots & \ddots & \vdots \\
\mu_{i_t}(\beta_\rho\beta_1^{s_l})  & \mu_{i_t}(\beta_\rho\beta_2^{s_l})  & \ldots & \mu_{i_t}(\beta_\rho\beta_m^{s_l})  \\
\end{array}\right] \cdot 
\left[ \begin{array} {c}
\mu_1(\bg(\gamma)) \\
\mu_2(\bg(\gamma)) \\
\vdots \\
\mu_m(\bg(\gamma))
\end{array} \right]  \label{connect-proof}
\end{equation}
where the matrix on the right side is exactly a folded row of ${\boldsymbol \Gamma}$.
The above equality immediately indicates that ${\boldsymbol \Gamma}$ corresponds to  the coefficient vector of null-subspace of the expanded generator $\bG_e$.
 Therefore, if a linear combination of rows of ${\boldsymbol \Gamma}$ results in an all-zero row, then the linear combination with respect to $\bG_e$
results in a valid subspace codeword. On the other hand, following Lemma~\ref{LMoneElement}, 
a valid subspace codeword exhibits the all-zero coefficient vector associated with null-subspace. 
The lower bound is obtained by assuming the worst-case that  ${\boldsymbol \Gamma}$ is full-rank.\\
$(ii)$. Theorem~\ref{TH-single-term} indicates that 
\begin{eqnarray*}
\aline{}\left[\mu_1(\bg(\gamma)),\; \mu_2(\bg(\gamma)),\; \ldots,\; \mu_m(\bg(\gamma))\right] \\
\aline{=} \left[{\boldsymbol 0}, \ldots, {\boldsymbol 0}, \mu_{j_1}(\bg(\gamma)), {\boldsymbol 0}, \ldots, {\boldsymbol 0},  
\mu_{j_2}(\bg(\gamma)), {\boldsymbol 0}, \ldots, {\boldsymbol 0}, \mu_{j_r}(\bg(\gamma)), {\boldsymbol 0}, \ldots, {\boldsymbol 0} \right].
\end{eqnarray*}
 Thus, the equality \eqref{connect-proof} is reduced to
 $$\left[\begin{array} {c}
\mu_{i_1}(\beta_\rho\bg(\gamma^{q^{s_l}})) \\
 \mu_{i_2}(\beta_\rho \bg(\gamma^{q^{s_l}})) \\
\vdots \\
 \mu_{i_t}(\beta_\rho \bg(\gamma^{q^{s_l}}))
\end{array}\right]
=\left[\begin{array}{c c c c}
\mu_{i_1}(\beta_\rho\beta_{j_1}^{s_l})  & \mu_{i_1}(\beta_\rho\beta_{j_2}^{s_l})  & \ldots & \mu_{i_1}(\beta_\rho\beta_{j_r}^{s_l})  \\
\mu_{i_2}(\beta_\rho\beta_{j_1}^{s_l})  & \mu_{i_2}(\beta_\rho\beta_{j_2}^{s_l})  & \ldots & \mu_{i_2}(\beta_\rho\beta_{j_r}^{s_l})  \\
\vdots & \vdots & \ddots & \vdots \\
\mu_{i_t}(\beta_\rho\beta_{j_1}^{s_l})  & \mu_{i_t}(\beta_\rho\beta_{j_2}^{s_l})  & \ldots & \mu_{i_t}(\beta_\rho\beta_{j_r}^{s_l})  \\
\end{array}\right] \cdot 
\left[ \begin{array} {c}
\mu_{j_1}(\bg(\gamma)) \\
\mu_{j_2}(\bg(\gamma)) \\
\vdots \\
\mu_{j_r}(\bg(\gamma))
\end{array} \right]
$$
which concludes the part $(ii)$. \hfill $\Box\Box$

Next theorem shows a complementary view on the dimension of the subspace subcodes.
\begin{theorem}   \label{TH-alt-dim}
Given  an expanded generator matrix 
$\bG_e( \gamma, \gamma^{q^{s_1}}$, $\gamma^{q^{s_2}}$,  \ldots,   $\gamma^{q^{s_{k-1}}} )$, 
the dimension of the subspace subcode with respect to a subbasis 
$\{\beta_{i_1}, \beta_{i_2}, \ldots, \beta_{i_t}\}$ is 
\begin{equation}
m_\gamma \cdot (t-{\mathcal R}({\boldsymbol \Theta}) ),
\end{equation}
where ${\mathcal R}({\boldsymbol \Theta})$ denotes the maximum number of linearly independent (defined in $\GF(q^{m_\gamma})$) rows 
of ${\boldsymbol \Theta}$ defined as
\begin{equation}
{\boldsymbol \Theta} \eqdef 
\left[ \begin{array}{cc c c}
\beta_{i_1}^{q^{m_\gamma-z_1}} & \beta_{i_1}^{q^{m_\gamma-z_2}} & \ldots &  \beta_{i_1}^{q^{m_\gamma-z_\kappa}} \\
\beta_{i_2}^{q^{m_\gamma-z_1}} & \beta_{i_2}^{q^{m_\gamma-z_2}} & \ldots &  \beta_{i_2}^{q^{m_\gamma-z_\kappa}} \\
\vdots & \vdots & \ddots & \vdots \\
\beta_{i_t}^{q^{m_\gamma-z_1}} & \beta_{i_t}^{q^{m_\gamma-z_2}} & \ldots & \beta_{i_t}^{q^{m_\gamma-z_\kappa}} 
\end{array} \right]
\end{equation}
where $\kappa\eqdef m_\gamma -k$ and 
\begin{equation}
\{{z_1}, {z_2}, \ldots, {z_\kappa} \} \eqdef
\{1, 2, \ldots, m_\gamma-1 \} \big\backslash\, \{ {s_1}, s_2, \ldots, s_{k-1} \}.  \label{def-power-zs}
\end{equation}
\end{theorem}

{\em Proof: }
It is easily seen that any $\chi_{\gamma}(x)$, $\chi_{\gamma^{q^{s_1}}}(x)$, \ldots, 
$\chi_{\gamma^{q^{s_{k-1}}}}(x)$, 
where the function $\chi$ is defined in \eqref{def-chi}, share $\kappa=m_\gamma-k$ common conjugacy roots, 
$\{\gamma^{-q^{z_1}}, \gamma^{-q^{z_2}}, \ldots, \gamma^{-q^{z_\kappa}} \}$.
We next define the polynomial
$$P(x)\eqdef \beta_{i_1}P_1(x)+\beta_{i_2}P_2(x)+\ldots+\beta_{i_t}P_t(x),$$
where $P_i(x)\in \GF(q)[x]$, $\deg(P_i(x))<m_\gamma$, $i=1, 2, \ldots, t$, and $P(\gamma^{-q^{z_l}})=0$, $l=1, 2, \ldots, \kappa$.

We proceed to show the one-to-one map between a valid codeword polynomial and a polynomial
$P(x)$ defined as above. 
Note that a valid codeword polynomial $c(x)$ can be represented by 
\begin{eqnarray*}
c(x)\aline{=}\theta_0 g(\gamma) + \theta_1 g(\gamma^{q^{s_1}})+\ldots + \theta_{k-1} g(\gamma^{q^{s_{k-1}}}) \\
\aline{=} \frac{ x^{N}-1} {p_{\gamma^{-1}} (x) }  \left(  \theta_0 \chi_\gamma(x) + \theta_1 \chi_{\gamma^{q^{s_1}}} (x) + \ldots +
\theta_{k-1} \chi_{\gamma^{q^{s_{k-1}}}} (x) \right),
\end{eqnarray*}
where $\theta_i\in \GF(q^m)$, $i=0, 1, 2, \ldots, {k-1}$. Clearly, 
$$P(x)=\theta_0 \chi_\gamma(x) + \theta_1 \chi_{\gamma^{q^{s_1}}} (x) + \ldots +
\theta_{k-1} \chi_{\gamma^{q^{s_{k-1}}}} (x)$$ 
can be represented by the subbasis $\{\beta_{i_1}, \beta_{i_2}, \ldots, \beta_{i_t}\}$ and 
contains the roots $\gamma^{-q^{z_1}}$, $\gamma^{-q^{z_2}}$, \ldots, $\gamma^{-q^{z_\kappa}}$.
Conversely, note that a polynomial is a valid codeword polynomial if it contains the roots $\GF^*(q^m)\backslash$ 
$\{\gamma^{-1}$, $\gamma^{-q^{s_1}}$, \ldots, $\gamma^{-q^{s_{k-1}}} \}$ and can be represented by the subbasis 
$\{\beta_{i_1}$, $\beta_{i_2}$, \ldots, $\beta_{i_t}\}$.
Clearly, $\frac{ x^{N}-1} {p_{\gamma^{-1}} (x) } P(x)$ is a valid codeword polynomial.

The condition that $P(x)$ contains the roots $\gamma^{-q^{z_1}}, \gamma^{-q^{z_2}}, \ldots, \gamma^{-q^{z_\kappa}}$ indicates
\begin{equation*}
\left\{\begin{array}{l c l}
0\aline{=}\beta_{i_1} P_1(\gamma^{-q^{z_1}}) + \beta_{i_2} P_2(\gamma^{-q^{z_1}}) +\ldots+ \beta_{i_t} P_t(\gamma^{-q^{z_1}}) \\
0\aline{=}\beta_{i_1} P_1(\gamma^{-q^{z_2}}) + \beta_{i_2} P_2(\gamma^{-q^{z_2}}) +\ldots+ \beta_{i_t} P_t(\gamma^{-q^{z_2}}) \\
\aline{\vdots}  \\
0\aline{=}\beta_{i_1} P_1(\gamma^{-q^{z_\kappa}}) + \beta_{i_2} P_2(\gamma^{-q^{z_\kappa}}) +\ldots+ \beta_{i_t} P_t(\gamma^{-q^{z_\kappa}})
\end{array}\right.  
\end{equation*}
We observe that
\begin{eqnarray*}
0 \aline{=} 
\left( \beta_{i_1} P_1(\gamma^{-q^{z_l}}) + \beta_{i_2} P_2(\gamma^{-q^{z_l}}) +\ldots+ \beta_{i_t} P_t(\gamma^{-q^{z_l}}) \right)^{q^{m_\gamma-z_l}} \\
\aline{=}\beta_{i_1}^{q^{m_\gamma-z_l}} P_1(\gamma^{-1}) + \beta_{i_2}^{q^{m_\gamma-z_l}} P_2(\gamma^{-1}) + \ldots + \beta_{i_t}^{q^{m_\gamma-z_l}} P_{t}(\gamma^{-1}).
\end{eqnarray*}
Therefore, the preceding equation system can be transformed into
\begin{equation*}
\left[ \begin{array}{cc cc cc cc}
\beta_{i_1}^{q^{m_\gamma-z_1}} & \beta_{i_2}^{q^{m_\gamma-z_1}} & \ldots &  \beta_{i_t}^{q^{m_\gamma-z_1}}  \\
\beta_{i_1}^{q^{m_\gamma-z_2}} & \beta_{i_2}^{q^{m_\gamma-z_2}} & \ldots &  \beta_{i_t}^{q^{m_\gamma-z_2}}  \\
\vdots  & \vdots  & \ddots & \vdots \\
\beta_{i_1}^{q^{m_\gamma-z_\kappa}} & \beta_{i_2}^{q^{m_\gamma-z_\kappa}} & \ldots &  \beta_{i_t}^{q^{m_\gamma-z_\kappa}} 
\end{array} \right]
\left[ \begin{array}{c}
P_1 \\ P_2 \\ \vdots \\ P_t
\end{array} \right]
=\left[ \begin{array}{c}
0 \\ 0 \\ \vdots \\ 0
\end{array} \right]
\end{equation*}
where $P_i$ denotes $P_i( \gamma^{-1} )$. It follows that the dimension of the solution space
is determined by the matrix ${\boldsymbol \Theta}$. 
Further note that $p_i=P_i(\gamma^{-1})$  where $p_i\in \GF(q^{m_\gamma})$ and $\deg(P_i(x))<m_\gamma$  uniquely 
determines the polynomial $P_i(x)\in\GF(q)[x]$. Finally, due to the homogeneousness of the above system, 
each indepedent solution $\{p_1$, $p_2$, \ldots, $p_t\}$ can be arbitrarily scaled within $\GF(q^{m_\gamma})$, and  
 thus exhibits a dimension of $m_\gamma$. Finally, it is worth noting that the linear dependence must be in light of $\GF(q^{m_\gamma})$,  
because 
$P_i(\gamma) \in \GF(q^{m_\gamma})$.
The theorem follows. \hfill $\Box\Box$

\begin{corollary}  \label{COR-conjugacy}
Given  an expanded generator matrix 
$\bG_e( \gamma, \gamma^{q^{s_1}}$, $\gamma^{q^{s_2}}$,  \ldots,   $\gamma^{q^{s_{k-1}}} )$,\\
$(i)$. there exist codewords with respect to a subbasis of $m-k+1$ elements.\\
$(ii)$. when $k=m-1$, no codeword can be represented by a single-element subbasis;
when $k=m-2$, there exist codewords to be represented by a two-element subbasis $\{\beta_1, \beta_2\}$ if and only if
$\beta\eqdef \frac{\beta_2}{\beta_1}$ is a sub-field element and furthermore, $m_\beta$ divides $|z_2-z_1|$,
where $z_1, z_2$ are as defined in \eqref{def-power-zs}.\\
$(iii)$. if a codeword $\bc$ is represented by a minimal subbasis of $i$ elements 
(herein ``minimal'' means that any proper subset fails), 
then its weight is equal to $ i\cdot q^{m_\gamma-1}(q-1) \frac{q^m-1}{q^{m_\gamma}-1} $.
\end{corollary}
{\em Proof:} Part $(i)$ is due to that the number of rows of ${\mathcal \Gamma}$ is $m(k-1)$, and thus results in a positive dimension value.
Part $(ii)$ straightforwarly follows Theorem~\ref{TH-alt-dim} (for the assertion of $k=m-2$, it comes down to 
$\beta^{q^{|z_2-z_1|}}=\beta$). \\
$(iii)$. We observe that the polynomial $\frac{x^{N}-1}{p_{\gamma^{-1}}(x)}$ is 
a common factor of $\{g_{\gamma^{q^j}}(x):\; 0\leq j< m\}$.
Thus, $\mu_l(\beta_i \bg(\gamma^{q^j}))$, $1\leq l, j, i\leq m$  are codewords associated with 
the generator polynomial $\frac{x^{N}-1}{p_{\gamma^{-1}}(x)}$. 
This also holds true for {\bf y} that is a linear combination of the conjugacy set 
$\{\beta_i\bg(\gamma^{q^j}): \; 1\leq i\leq m, 0\leq j <m\}$.
Therefore, the conclusion follows Theorem~\ref{TH-equalWt}.
\hfill $\Box\Box$

Note in the extreme case where $\gamma_1$, $\gamma_2$, \ldots, $\gamma_k$ compose a complete conjugacy class, 
$\bG_e(\gamma_1$, $\gamma_2$, \ldots, $\gamma_k)$
corresponds to a BCH subcode, as explored in \cite{Vardy}.
We now present examples in $\GF(2^8)$  to clarify the above theorem. $\bg(\alpha^1)$ and $\bg(\alpha^2)$ do not belong to the conjugacy class of 
a nontrivial subfield, and thus can be combined in a way to produce 
eight linearly independent binary codewords which are represented by a subbasis with seven elements; 
whereas $\bg(\alpha^1)$ and $\bg(\alpha^{16})$ compose the conjugacy class of 
the subfield $\GF(2^4)$, and thus can be combined in a way,  under an appropriate basis (say a composite basis 
$\{1$, $\alpha^{17}$, $\alpha^{34}$, $\alpha^{51}$, $\alpha^{1}$, $\alpha^{18}$, $\alpha^{35}$, $\alpha^{52}\}$), to produce 
eight linearly independent binary codewords which are represented by a subbasis of four elements 
(herein $\{1$, $\alpha^{17}$, $\alpha^{34}$, $\alpha^{51}\}$, or $\{\alpha^{1}$, $\alpha^{18}$, $\alpha^{35}$, $\alpha^{52}\}$).  
$\bg(\alpha^1)$, $\bg(\alpha^2)$ and $\bg(\alpha^{4})$ may be combined to produce codewords that are represented by a subbasis with six elements; 
$\bg(\alpha^1)$, $\bg(\alpha^2)$ and $\bg(\alpha^{16})$ may be combined to produce codewords to be represented by a subbasis with four elements;
$\bg(\alpha^1)$, $\bg(\alpha^4)$, $\bg(\alpha^{16})$, and $\bg(\alpha^{64})$, under an appropriate basis (say a composite basis
$\{1$, $\alpha^{17}$, $\alpha^{85}$, $\alpha^{102}$, $\alpha^{1}$, $\alpha^{18}$, $\alpha^{86}$, $\alpha^{103}\}$), 
 may be combined to produce codewords that are represented by a subbasis with two elements 
(herein $\{1, \alpha^{85}\}$, or $\{\alpha^{17}, \alpha^{102}\}$, or $\{\alpha, \alpha^{86}\}$, or $\{\alpha^{18}, \alpha^{103}\}$).

We proceed to establish the (negative) relation between subbasis and non-conjugate elements.
\begin{theorem}  \label{THnonconjugacy}
Given an expanded generator matrix $\bG_e( \gamma_1, \gamma_2, \ldots, \gamma_k)$
where $\gamma_i$, $i=1$, 2, \ldots, $k$, are non-subfield elements and satisfy $\phi(\gamma_i) \ne \phi(\gamma_j), \;\forall i\ne j$,
no nontrivial codeword can be represented by a proper subbasis. 
\end{theorem}
{\em Proof:} We show the correctness by contradiction. Assume there exists a nonzero codeword 
$\sum_{j=1}^k \theta_j \bg(\gamma_j)$ (where $\theta_j\in \GF(q^m)$)
which can be represented by a proper subbasis, say $\beta_l$ not being included. 
It is shown in Theorem~\ref{TH-single-term} that the codeword with only one nontrivial coefficient $\theta_j$ can not be represented by a subbasis.
We proceed to consider the remaining cases where at least two coefficients are nontrivial.
Without loss of generality, we assume $\theta_1$ and $\theta_2$ are nontrivial.
Recall that (as shown in \eqref{component-g-x}) the polynomial 
$\mu_l\left( \theta_1 g_{\gamma_1}(x) \right)$ 
is not divisible by $p_{\gamma^{-1}_1}(x)$, whereas the all other polynomials 
$\mu_l\left( \theta_i g_{\gamma_i}(x) \right)$, $i=2, 3, \ldots k$, are all divisible by $p_{\gamma^{-1}_1}(x)$.
Therefore, $\theta_1$ must be trivial.  The theorem follows. \hfill $\Box\Box$ 

In \cite{Hattori}, a explicit formula utilizing dual-basis is given for determining the dimension of subspace subcodes defined in $\GF(2^m)$. 
Clearly, Theorems \ref{THconjugacy}, \ref{TH-alt-dim} and \ref{THnonconjugacy} reveal an alternative and more general interpretation 
on the dimension of subspace subcodes, and particularly reveal that 
the dimension of a supspace subcode can be optimized through choosing appropriate composite basis.
We next present examples to clarify the above assertion. When a 6-dimensional subspace in $\GF(2^8)$ is considered, we maximize the subcode dimension by employing the composite basis 
$\{1, \alpha^{17}, \alpha^{85}, \alpha^{102}, \alpha^1, \alpha^{18}, \alpha^{86}, \alpha^{103} \}$ and subsequently choosing the subbasis 
$\{1, \alpha^{17}, \alpha^{85}, \alpha^{102}, \alpha^1, \alpha^{86} \}$. 
Under this subspace, $\bG_e(\alpha, \alpha^4)$ contains a subcode of eight dimensions 
(recall that $\alpha$ and $\alpha^4$ pertain to the same conjugacy class of the subfield $\GF(2^2)$), whereas under a regular polynomial basis 
it doesnot contain a 6-dimensional subspace subcode. 
$\bG_e(\alpha^{17})$ contains a subcode with dimension 4 (recall that $\bg(\alpha^{17})$, $\alpha^{17}\bg(\alpha^{17})$, $\alpha^{85}\bg(\alpha^{17})$, 
$\alpha^{102} \bg(\alpha^{17})$ can be represented by the subbasis $\{1, \alpha^{17}, \alpha^{85}, \alpha^{102}\}$. 
$\bG_e(\alpha^{85})$ contains a subcode with dimension 6 
(recall that $\bg(\alpha^{85})$ and $\alpha^{85}\bg(\alpha^{85})$ can be represented by the subbasis $\{1, \alpha^{85}\}$;
$\alpha^{17}\bg(\alpha^{85})$ and $\alpha^{102}\bg(\alpha^{85})$ can be represented by the subbasis $\{\alpha^{17}, \alpha^{102}\}$;
$\alpha\bg(\alpha^{85})$ and $\alpha^{86}\bg(\alpha^{85})$ can be represented by the subbasis $\{\alpha, \alpha^{86}\}$).

The following corollary is an extension of Corollary~\ref{CO-common-poly}.
\begin{corollary}
Given an expanded generator matrix $\bG_e(\gamma_1$, $\gamma_2$, \ldots, $\gamma_k)$, 
let $l$ be the minimum number of basis elements to represent any particular codeword generated by 
$\bG_e(\phi(\gamma) \cap \{\gamma_1, \gamma_2, \ldots, \gamma_k\})$,
 then $p_{\gamma^{-1}}(x)$ divides either all $m$ component polynomials or up to $m-l$ component polynomials of any codeword polynomial generated by 
$\bG_e(\gamma_1$, $\gamma_2$, \ldots, $\gamma_k)$.
\end{corollary}
{\em Proof: } If a codeword is generated by $\bG_e(\{\gamma_1, \gamma_2, \ldots, \gamma_k\}\backslash \phi(\gamma) )$,
then $p_{\gamma^{-1}}(x)$ divides all $m$ component polynomials, as shown in Corollary~\ref{CO-common-poly}. Otherwise, we divides the codeword $\bc$ into two parts
$\bc=\bc_1+\bc_2$, where $\bc_1$ is generated by $\bG_e(\{\gamma_1, \gamma_2, \ldots, \gamma_k\}\backslash \phi(\gamma) )$ and 
$\bc_2$ is generated by  $\bG_e(\phi(\gamma) \cap \{\gamma_1, \gamma_2, \ldots, \gamma_k\})$. 
We note that $p_{\gamma^{-1}}(x)$ divides all component codeword polynomials of
$\bc_1$, while divides only the zero component polynomials of $\bc_2$. The corollary follows. \hfill $\Box\Box$

The following corollary characterizes the number of linearly independent components of a codeword.
\begin{corollary}
Let a codeword $\bc$ be composed of
$$\bc=\theta_1\bg(\gamma_1) + \theta_2\bg(\gamma_2) + \ldots+ \theta_k\bg(\gamma_k), \hspace{0.2in} \theta_i\ne 0,\; \forall\; i,$$
and $l_i$ be the minimum size of subbasis to represent any particular codeword generated by $\bG_e(\phi(\gamma_i)\cap\{\gamma_1, \gamma_2, 
\ldots, \gamma_k\})$ under a given basis $\{\beta_1, \beta_2, \ldots, \beta_m\}$. 
Then, the number of linearly independent component codewords of $\bc$ is at least $\max\{l_1, l_2, \ldots, l_k\}$. 
\end{corollary}
{\em Proof: } Without loss of generality, let $l_1$ be the largest, i.e., $l_1\geq l_i$, $i=2$, 3, \ldots, $k$.
We first decompose the codeword $\bc$ into two parts, $\bc=\bc_1+\bc_2$, such that $\bc_1$ corresponds to the generator matrix 
$\bG_e(\phi(\gamma_1)\cap\{\gamma_1, \gamma_2, \ldots, \gamma_k\})$, and $\bc_2$ corresponds to the generator matrix 
$\bG_e(\{\gamma_1, \gamma_2, \ldots, \gamma_k\}\backslash\phi(\gamma_1))$.
We first show that the number of linearly independent component codewords of $\bc_1$ is at least $l_1$  through contradiction. 
Without loss of generality, we assume that the component codewords,  $\mu_{l}(\bc_1)$, $\mu_{l+1}(\bc_1)$, \ldots, $\mu_{m}(\bc_1)$, are 
linearly dependent on the linearly independent component codewords, $\mu_1(\bc_1)$, $\mu_2(\bc_1)$, \ldots, $\mu_{l-1}(\bc_1)$ ($l\leq l_1$), 
such that,
$$\mu_i(\bc_1) = \nu_{i, 1} \mu_1(\bc_1) + \nu_{i, 2} \mu_2(\bc_1) +\ldots + \nu_{i, l-1} \mu_{l-1}(\bc_1),
\hspace{0.2in} i=l, l+1, \ldots, m.$$
Consequently, we obtain
\begin{eqnarray*}
\bc_1 \aline{=} \sum_{i=1}^{l-1}\beta_i \mu_i(\bc_1) + \sum_{i=l}^{m}\beta_i \mu_i(\bc_1) \\
\aline{=}  \sum_{i=1}^{l-1}\beta_i \mu_i(\bc_1) + \sum_{i=l}^{m}\beta_i  \sum_{j=1}^{l-1} \nu_{i, j} \mu_j(\bc_1) \\
\aline{=} \sum_{i=1}^{l-1}  \mu_i(\bc_1) (\beta_i+ \sum_{j=l}^{m} \nu_{j, i} \beta_{j}). 
\end{eqnarray*}
The above equality indicates that
 $\bc_1$ can be represented by the subbasis
$$\left\{  \beta_1+\sum_{i=l}^{m} \nu_{i, 1} \beta_{i}, \;\beta_2+\sum_{i=l}^{m} \nu_{i, 2} \beta_{i}, \; \ldots, \;
\beta_{l-1}+\sum_{i=l}^{m} \nu_{i, l-1} \beta_{i} \right\},$$
which has $l-1\leq l_1-1$ elements (and can be expanded to an alternative basis). This clearly violates  the definition of $l_1$. 

On the other hand, we recall that $p_{\gamma^{-1}_1}(x)$ divides all component polynomials of $\bc_2$, whereas it divides only all-zero component polynomials of 
$\bc_1$. Therefore, adding $\bc_2$ to $\bc_1$ cannot reduce the number of linearly independent component codewords of $\bc_1$.
We thus conclude the corollary. \hfill $\Box\Box$

In \cite{Sak}, a lower bound on the minimum distance of expanded cyclic codes is obtained by treating it as a generalized concatenated code. 
The following theorem establishes an improved bound by incorporating the preceding new insights.
\begin{theorem}  \label{TH-dmin-bound}
Given an expanded generator matrix $\bG_e(\gamma_1$, $\gamma_2$, \ldots, $\gamma_k)$, the minimum distance is bounded by 
\begin{equation}
\dmin \geq \min_{1\leq i\leq m} \{i\cdot d^{(i)}\},
\end{equation}
where $d^{(i)}$ denotes the minimum distance of the subcode associated with the generator polynomial
\begin{equation}
G^{(i)}(x)=\frac{x^{N}-1}{\text{LCM}\{p_{\gamma^{-1}_i}(x):\; \phi(\gamma_i)\cap \{\gamma_1, \gamma_2, \ldots, \gamma_k\} 
\text{ results in a subbasis with up to } i \text{ elements}\} }
\end{equation}
where LCM stands for ``Least Common Multiplier".
\end{theorem}
In essence, in \cite{Sak}, the minimum distance of an outer code is shown to be bounded by the largest number
of consecutive conjugate elements, whereas it is precisely computed through Theorems~\ref{THconjugacy}, \ref{TH-alt-dim}, \ref{THnonconjugacy}. 

We present three examples in $\GF(2^5)$ to shed light on the proposed bound in contrast to the bound in \cite{Sak}. 
Given the generator matrix 
$\bG_e(\alpha^{21}, \alpha^{22})$, where $\alpha^{22}=\alpha^{21\times 4}$,
the proposed lower bound is computed as $16\times 4=64,$ whereas the bound provided in \cite{Sak} is 48.
Given the generator matrix 
$\bG_e(\alpha^{21}, \alpha^{22}, \alpha^{23})$, 
the proposed lower bound is $\min\{16\times 4,\; 12\times 5\}=60,$ whereas the bound provided in \cite{Sak} is 48.
Given the generator matrix 
$\bG_e(\alpha^{18}, \alpha^{19}, \alpha^{20}, \alpha^{21}, \alpha^{22})$, where $\alpha^{20}=\alpha^{18\times 8}$ and 
$\alpha^{22}=\alpha^{21\times 4}$,
the proposed lower bound is $\min\{10\times 4, \; 8\times 5\}=40,$ whereas the bound provided in \cite{Sak} is 36.

It is worth noting that the proposed bound is rather loose for high rate codes. 
For instance, let the code rate of a Reed-Solomon code in $\GF(2^m)$ be greater than one half, then,
when $m$ is a prime, the proposed bound on the minimum distance of the resulting expanded code reduces trivially to $2m$, as $G^{(m)}(x)=x-1$ and subsequently $d^{(m)}=2$ 
(actually the worse case is that $G^{(m)}(x)=1$ and $d^{(m)}=1$); 
alternatively, when $m$ is not a prime, $G^{(m)}(x)$ may contain the minimal polynomials of subfield elements, 
and thus the bound can be somewhat improved.

The following theorem shows that the binary expanded Reed-Solomon codes, regardless of realization basis, are asymptotically bad, 
in contrary to the prevalent conjecture (cf. \cite{El-Khamy}), as well as to the ensemble of generalized Reed-Solomon codes 
which asymptotically achieves the Gilbert-Varshamov bound \cite{Retter}. 

\begin{theorem}
For a sequence of primitive $(2^m-1, \lfloor(2^m-1)\rfloor )$ Reed-Solomon codes with a fixed rate $r$ and a fixed starting spectrum $\delta$ 
(i.e., its generator polynomial is defined as $G(x)=\prod_{i=\delta}^{\delta-1+\lceil(1-r)(2^m-1)\rceil} (x-\alpha^{i})$), 
their binary Hamming minimum distances $d_{\min}$ satisfy
\begin{equation}
\lim_{m\rightarrow \infty} \frac{d_{\min}}{m(2^m-1)}=0.
\end{equation} 
\end{theorem}
{\em Proof:} We first consider the case $\delta \geq 0 $. Let 
$$k=\lfloor \log_2 ((2^m-1)r-\delta) \rfloor . $$
We observe that the set $\{\beta_i\bg(\alpha^{q^s}):\; 1\leq i\leq m,\; 0\leq s \leq k \}$ 
is contained in the binary expansion of the generator matrix.
In accordance with Theorem~\ref{THconjugacy}, there exists a codeword 
that is represented by a subbasis with up to $m-k$ elements and is with weight at most  $(m-k)2^{m-1}$.
Therefore,
\begin{eqnarray*}
\lim_{m\rightarrow \infty} \frac{d_{\min}}{m(2^m-1)} 
\aline{\leq} \lim_{m\rightarrow \infty} \frac{(m-k)2^{m-1}}{m(2^m-1)} \\
\aline{\leq} \lim_{m\rightarrow \infty} \frac{(m- \log_2 (r2^m-r-\delta) +1 )2^{m-1}} {m(2^m-1)} \\
\aline{=}  0.
\end{eqnarray*}
Now we consider the alternative case  $\delta < 0 $. Let
\begin{eqnarray*}
k_1\aline{=}\lceil \log_2 (-\delta) \rceil, \\
k_2\aline{=}\lfloor \log_2 ((2^m-1)r-\delta) \rfloor . 
\end{eqnarray*}
Following Corollary~\ref{COR-conjugacy}.$(i)$, there exists a codeword 
that is represented by a subbasis with up to $m-(k_2+k_1)$ elements and is with weight at most $(m-k_2+k_1)2^{m-1}$.
Again, we have 
\begin{eqnarray*}
\lim_{m\rightarrow \infty} \frac{d_{\min}}{m(2^m-1)} 
\aline{\leq} \lim_{m\rightarrow \infty} \frac{(m-k_2+k_1  )2^{m-1}}{m(2^m-1)} \\
\aline{\leq} \lim_{m\rightarrow \infty} \frac{(m- \log_2 (r2^m-r-\delta) + \lceil \log_2 (-\delta) \rceil +1 )2^{m-1}} {m(2^m-1)} \\
\aline{=}  0.
\end{eqnarray*}
The proof is completed. \hfill $\Box\Box$

In addition, elements of very small subfield also contribute to low weight. E.g., when $m$ is even,
$$w\left(\bg(\alpha^{N/3})\right)=2\times 2^{2-1} \frac{2^m-1}{2^2-1}=\frac{4N}{3}$$
under an appropriate composite basis 
(say $\{1, \alpha^1, \ldots, \alpha^{m/2-1}, \alpha^{N/3}, \alpha^{N/3+1}, \ldots, \alpha^{N/3+m/2-1} \}$), 
where $\alpha^{N/3}$ pertains to the subfield $\GF(2^2)$.

\section{Concluding Remarks}

The paper has a threefold purpose. The first purpose is to present an explicit description of expanded cyclic codes defined in $\GF(q^m)$.
The second purpose of this paper is to identify a class of constant-weight cyclic codes which achieve the Plotkin bound.
The last purpose of the paper is to characterize expanded cyclic codes utilizing the proposed expanded generator matrix and parity check matrix. 
We characterize the properties of component codewords of a codeword and particularly identify the precise conditions 
under which a codeword can be represented by a subbasis.

Our analysis seems to suggest that symbol-wise minimum weight codewords are irrelevant to 
the bit-wise minimum weight codewords. Our extensive  simulations suggest that the component codewords corresponding to different 
indices may not reach (close to) minimum weight simultaneously and subsequently the proposed the minimum distance bound is rather loose 
(for instance, when the code rate of a Reed-Solomon code in $\GF(2^m)$ is greater than half, 
the proposed bound on the minimum distance of the resulting expanded code by and large reduces to $2m$).
Therefore, it is imperative to determine a substantially tighter bound. Moreover, we strongly believe that this is also critical to 
explicitly find ``good" codes from binary expanded cyclic codes (without generalization, which inevitably renders the analysis intractable).

\section*{Acknowledgement}

The author would like to thank Dr. Jun Ma, Prof. Jorn Justesen, and particularly, Prof. Marc Fossorier, for many constructive comments on improving the presentation of the manuscript.

\renewcommand{\baselinestretch}{1.0}

\end{document}